# ECG beat classification using machine learning and pre-trained convolutional neural networks

Neville D. Gai, PhD

*Systems Biology Center*

*National Heart, Lung, and Blood Institute*

*NIH, Bethesda, MD USA*

## Abstract

The electrocardiogram (ECG) is routinely used in hospitals to analyze cardiovascular status and health of an individual. Abnormal heart rhythms can be a precursor to more serious conditions including sudden cardiac death. Classifying abnormal rhythms is a laborious process prone to error. Therefore, tools that perform automated classification with high accuracy are highly desirable. The work presented classifies five different types of ECG arrhythmia based on AAMI EC57 standard and using the MIT-BIH data set. These include non-ectopic (normal), supraventricular, ventricular, fusion, and unknown beat. By appropriately transforming pre-processed ECG waveforms into a rich feature space along with appropriate post-processing and utilizing deep convolutional neural networks post fine-tuning and hyperparameter selection, it is shown that highly accurate classification for the five waveform types can be obtained. Performance on the test set indicated higher overall accuracy (98.62%), as well as better performance in classifying each of the five waveforms than hitherto reported in literature.

## Introduction

Electrocardiogram (ECG) provides a relevant window into the health of a person's cardiovascular system. Cardiovascular disease accounts for one-third of all deaths across the globe. In the US, heart disease is the leading cause of death in men and women. Therefore, correct diagnosis of cardiovascular disease is important. About one in four adults over 40 yrs old experience an irregular heartbeat. Arrythmias that last longer than a few seconds can result in sudden cardiac death (SCD). It is estimated that SCD from arrhythmia accounts for 15-20% of all deaths in the US. ECG based heartbeat classification assigns ECG to five different classes based on the Association for Advancement of Medical Instrumentation (AAMI) EC57 standard [1]. The five classes include normal (N), supraventricular (S), ventricular (V), fusion (F) and beats of unknown etiology (Q). The last four of these classes result in irregular heartbeats (arrythmias) which can be a serious condition resulting in sudden death. Supraventricular arrythmias originate in the region just above the ventricles or in the atria; ventricular arrythmias, as the name suggests, originate in the ventricles and result in greater fatality than other forms of arrythmias; fusion beats occur when supraventricular and ventricular impulses coincide to create a hybrid ECG waveform; arrythmias of unknown etiology usually comprise of a fusion of paced and normal beats and unclassifiable beats. Manually reading and performing the classification of arrhythmias is a laborious task and prone to errors. Therefore, predicting arrythmia class using machine learning is a worthy endeavor.

Recording and/or classifying ECG waveforms is a very active area of research. Several mobile devices along with proprietary algorithms are now available. Apple iWatch and companies like AliveCor's KardiaMobile with Kardia App which interface with smart watches have revolutionized mobile ECG detection and classification. However, they are mostly geared towards detecting atrial fibrillation (AF) rather than classifying beats into five classes as described above. Performance for such devices and algorithms is not readily available. In one study, AliveCor's automated algorithm provided 96.6% sensitivity and 94.1% specificity for AF while discarding 27.6% of unclassified beats [2]. A very recent version of the KardiaMobile device (KardiaMobile 6L) uses six leads and classifies beats into premature ventricular contractions, supraventricular ectopy and wide QRS, but exact performance numbers are not available. Information released by AliveCor regarding its newer AI classification algorithm (AI V2) indicates that unclassified beats are reduced by 80% while sensitivity and specificity for AF are > 90%. Other recent classification related literature includes Manta ray foraging optimization (MRFO) and SVM based algorithm [3], a method using 1D and 2D ECG signals fed to a combined recurrence plot (RP) and convolutional neural network (CNN) [4], a deep 2D convolution method to classify ECG data [5], a VGG based CNN [6], using GANs to augment the unbalanced data set along with CNN [7], combined CNN with Seq2Seq model [8], long short-term memory (LSTM) [9] and gated recurrent unit (GRU) [10].

Most machine learning methods for ECG processing usually follow three main steps, including a) signal pre-processing, which includes heartbeat segmentation, signal normalization, filtering etc. b) feature extraction; and c) modeling and learning. In this work, a novel framework for ECG analysis based on two improvements over prior works is presented: (a) feature extraction best suited for applying 2D CNN and (b) introducing spatial gradient across images obtained by transforming the 1D ECG waveform thus enabling effective application of CNN as discussed in Methods.

**Methods**

*Data Processing*:

PhysioNet's MIT-BIH ECG dataset [11] was used for training and classification. The dataset consists of 30 min ECG recordings of 48 individual subjects at a sampling frequency of 360 Hz. Each heartbeat has been annotated by two cardiologists (independently and by consensus in case of disagreement) to fall into one of five classes as described earlier — normal (N), supraventricular (S), ventricular (V), fusion (F) and unknown etiology (Q). While earlier works used a data split proposed by de Chazal et al. [12], more recent works have relied on data made available on Kaggle by Kachuee et al. [13]. This data set provides a pre-processed version sampled at 125 Hz from lead II (most relevant lead signal) to reduce data while capturing all relevant information. Briefly, data were split into 10 s windows and amplitudes normalized between 0 and 1. This is followed by determining maximums based on first derivative; the R-wave was then identified based on thresholding and median R-R time intervals were determined; beat period was calculated based on the identified R peak. A period of 1.2 s around peak was fixed with padding added if necessary. Typical beats along with their classes are shown in Figure 1. There are a total number of 109446 individual waveforms in the dataset which are divided in 80:20 ratio between training and testing sets such that the test set data is from different individuals than the training data. Number of examples available for training and testing in each of the five classes is shown in Table 1.

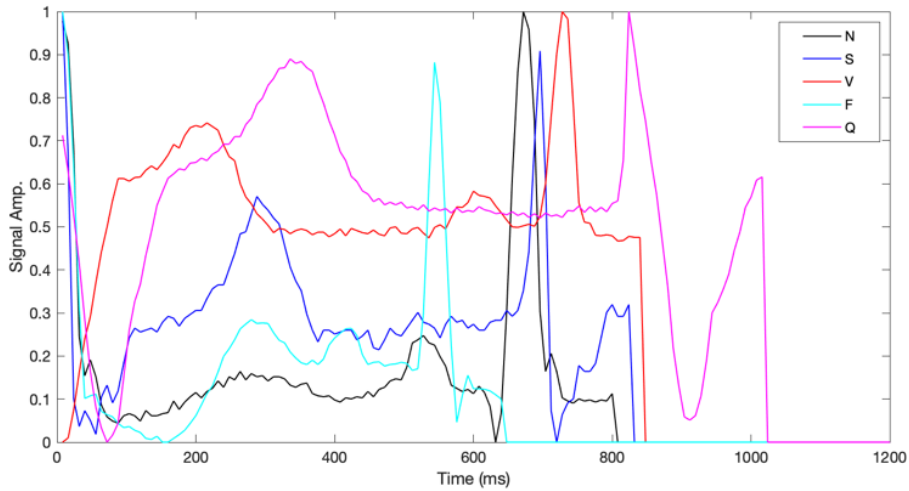

Figure 1: Exemplary ECG beats normalized to the R wave shown for N (normal), S (supraventricular), V (ventricular), F (fusion) and Q (unknown etiology) classes. Each ECG is sampled at 125 Hz and is of duration 1.5 s.

|  | F | N | Q | S | V |
|---|---|---|---|---|---|
| Training | 641 | 72471 | 6431 | 2223 | 5788 |
| Test | 162 | 18118 | 1608 | 556 | 1448 |

Table 1: Distribution of ECG examples for training and test sets.

*Feature Extraction*:

Feature extraction is an integral part of the machine learning (ML) approach. Transforming data into a rich feature space and applying effective models to the transformed data is a good strategy. Given the remarkable success achieved by 2D CNNs trained on the ImageNet data set, and the easy availability of several pre-trained 2D CNNs, transforming the 1D signal to a 2D image seemed a good strategy. The time-dependent ECG signal can be converted into a 2D signal in a variety of ways. Part of ML is to find a feature mapping which provides the best match to the problem. Based on literature survey, as well as the similarity of the 1D signal to radar signals and the well-developed field of radar signal processing, the Wigner-Ville distribution (WVD) was selected to transform the data. The WVD captures time and frequency variation in the signal ($s(t)$), effectively allowing a single comprehensive measure of the individual ECG signal. WVD is defined as

$$W_s(t, \omega) = \frac{1}{2\pi} \int s(t + \frac{\tau}{2}) s^*\left(t - \frac{\tau}{2}\right) e^{j\omega\tau} d\tau$$

where $\tau$ is the time shift. Among several interesting properties of WVD are that of being information loss-less, real valued along time and frequency, time and frequency co-variant, and dilation preserving. As shown in [14], the WVD should be used for classification of 1D non-stationary signals with 2D CNNs.

*Data processing*

1D ECG signals were sampled at 125 Hz before taking the WVD. An inherent problem with 2D CNNs is its translation invariant property. 2D images obtained by WVD of the 1D signals provide time encoded information which could be lost when applying a CNN. This issue was resolved by applying an appropriate gradient along the axis of the WVD image based on the work of Liu et al [15]. In the absence of heuristics related to the strength of gradient to be applied across each WVD image, several trials in combination with the CNN model provided an added value of 0.25 as near optimal. Figure 2 shows example images for the five classes.

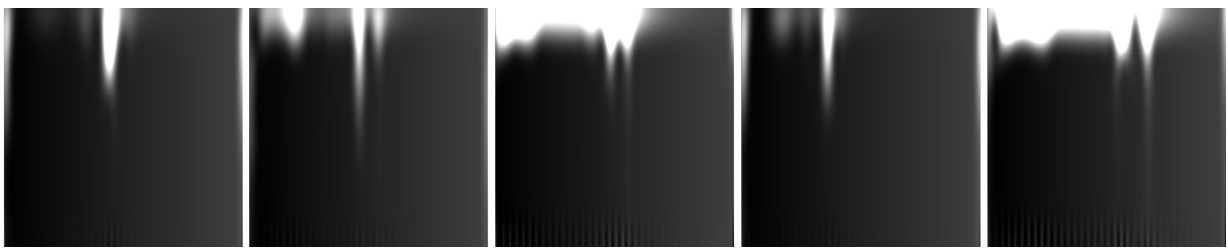

Figure 2: Exemplary WVD images of ECG beats corresponding (from left to right) to N, S, V, F and Q classes.

*CNN implementation*

Models pre-trained on ImageNet data set were considered for this study based on work by Ke et al. [16] which showed superior performance by ResNet and Inception CNNs, pretrained on ImageNet, for various medical imaging tasks. Initial implementation of the 2D CNN was done in Matlab® for ease of implementation. Data were split 80-20 between training and validation. Through trial-and-error, it was found that using ResNet18 with top and bottom layers modified worked best in Matlab®. The pre-trained ResNet18 network 71 layers deep was modified for the purpose of classifying ECG WVD images into one of five classes. The network is shown in Figure 3. Essential block of the ResNet consists of a convolution layer, followed by batch normalization and a *ReLu* activation layer. After several skipped blocks, the output from a previous *ReLu* layer is added to the output of stacks of the essential block forming a skipped layer connection. The input layer accepts images of size (128×128× 1); the convolution layer immediately following performs convolutions with 64 filters of size (7×7) and stride (2×2), so that the output of this layer has a size (64×64×64). The number of weights employed for this convolution layer is then (7×7×64) with 64 bias values corresponding to each of the 64 filters. The final output block consists of a ReLU layer, followed by a *max-pooling* layer, a fully connected (FC) layer with 5 outputs, a *softmax* layer and a final classification layer, classifying the WVD image into one of 5 classes based on *softmax* function probabilities. Input image size was (128,128,1) with classification into the 5 ECG arrythmia classes. Hyperparameters used were learning rate ($\eta$) = 0.01, regularization ($\lambda$) = 0.0001 with mini-batch gradient descent (batch size = 64) and epochs = 8.

To provide a more standardized and portable implementation, a similar approach was implemented in Keras using TensorFlow as backend on NVIDIA® Tesla® P100 GPU. A larger ResNet model (ResNet 50) was used. The input and output layers of the model were modified for the purpose of classifying into the five categories. A condensed layer graph for the final model along with number of training parameters is shown in Figure 4. To avoid conflict, all descriptions from here on forward will refer to the TensorFlow implementation, unless otherwise specified.

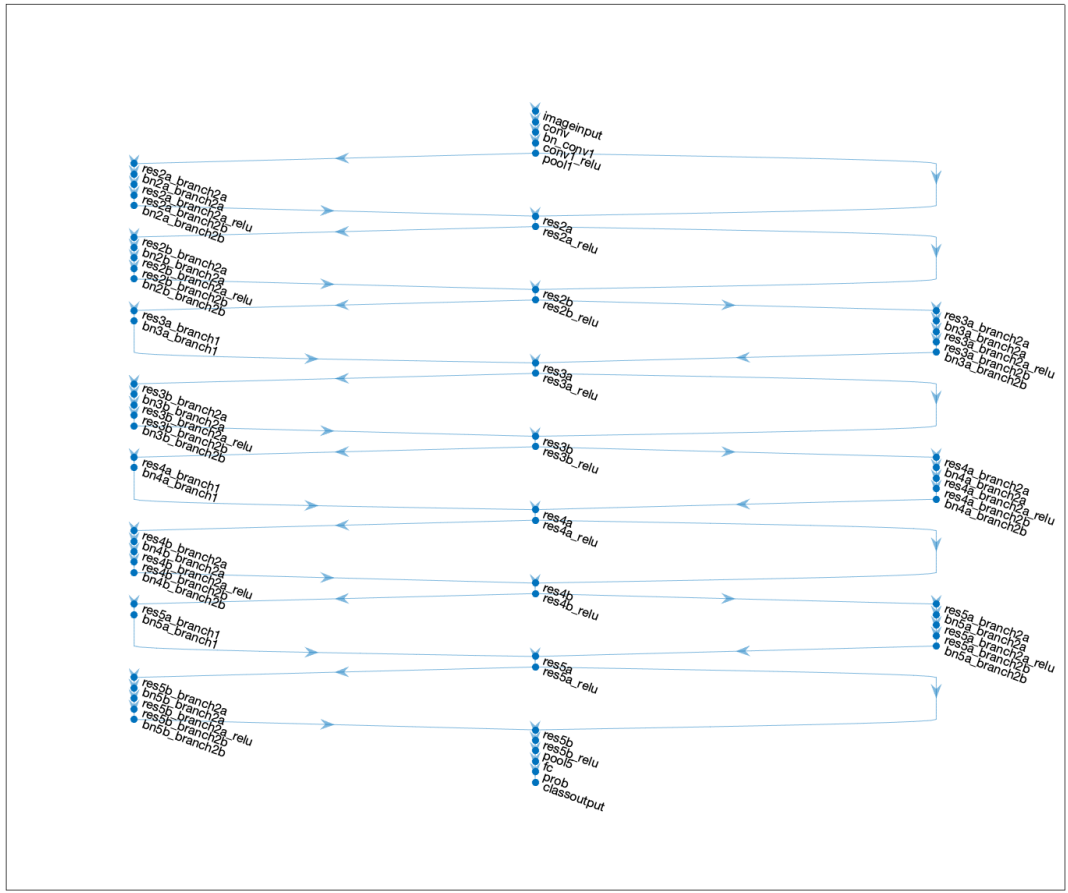

Figure 3: Layer graph of the ResNet18 model which was modified for the input and output of the ECG classification problem.

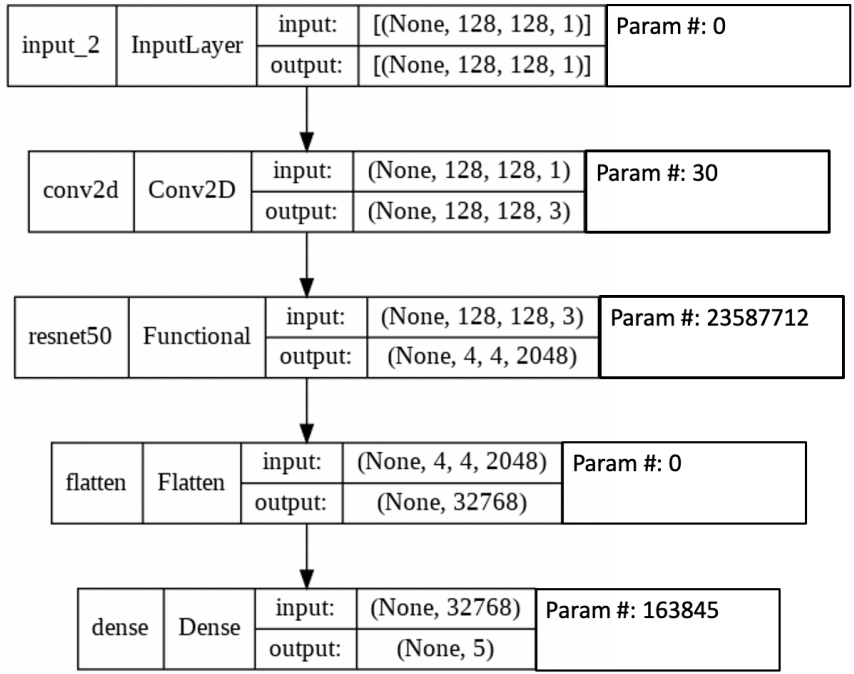

```
Total params: 23,751,587
Trainable params: 10,154,019
Non-trainable params: 13,597,568
```

Figure 4: Layer graph of the final selected model indicating layer names, input and output at each stage along with the number of trainable and non-trainable (due to freezing layers) parameters.

*Fine Tuning*

Three different layer configurations were considered for the pre-trained part of the network: (a) all trainable (b) first 125 frozen and (c) first 150 layers frozen. In addition, different configurations of the model were tried by concatenating additional layers after the flatten layer and prior to the *softmax* layer: (a) *Dense* layer with 1024 neurons, followed by *Dropout* (0.5) layer, followed by a *Dense* layer with 64 neurons (b) *Dense* layer with 64 neurons and (c) no additional layers.

*Hyperparameter selection*

For layers which could be regularized in the base model, L2 regularization was set with value $\lambda = 0.0001$. *Adam* optimizer with default values ($\beta_1 = 0.9$, $\beta_2 = 0.99$, $\varepsilon = 10^{-7}$) and mini-batch gradient descent were both tried. Different learning rates ($\eta = 0.001$, $\eta = 0.01$ and adaptive rate) with and without early stopping were considered. Mini-batch sizes of 32 and 64 were tried.

*Training*

The modified CNN model was implemented in TensorFlow version 2.7.0. Since differentiation was between five classes, *categorical crossentropy* was used as the loss function. Shuffling was employed with 20% of the data assigned for validation. Once reasonable hyperparameters were obtained based on performance over validation and test sets, optimal performance was always achieved by training the model over all training data without a validation set. Image normalization was not done since batch normalization follows convolution in the ResNet architecture. When early stopping was used, varying min($\Delta val\_acc$ or $\Delta acc$) and *patience* were employed.

*Data Augmentation*

Since the training data set is unbalanced, data augmentation of the minority sets was employed to check if classification would achieve better results for those sets. Flipping time varying ECG signals or translating and rotating them was not considered. Flipping images is not a good idea for this application, since it defeats the time dependent characteristic of the signal while translating and rotating would also change the nature of the signals in a class. Instead, as is common for such signals, Gaussian noise with maximum 10% of signal amplitude was added to create additional signals for minority classes. For N class with 72471 examples, a subset of 20000 examples was created by randomly picking them from the total. For a minority class like the S class with 2223 images, augmentation was done by adding Gaussian noise to (20000 – 2223 = 17777) ECG signals sampled in modulo 2223 fashion. WVD of the resulting waveforms provided a total of 20000 images per class. Another augmentation technique employed was to merely repeat the minority signals.

*Evaluation Metrics*

Precision, recall and F1 score were used as evaluation metrics since they capture all cases including true positive (TP), false positive (FP), true negative (TN) and false negative (FN).

$$Recall_i = \frac{TP_i}{TP_i + FN_i}; \ Precision_i = \frac{TP_i}{TP_i + FP_i}; \ F1score_i = \frac{2 \times Precision_i \times Recall_i}{Precision_i + Recall_i}$$

$TP_i$ refers to true positives for the $i^{th}$ class, $TN_i$ to true negatives, $FP_i$ to false positives and $FN_i$ to the false negatives. Recall, Precision and F1 score were calculated for each of the five classes.

The time for training the models is not as important as time to evaluate and classify a single beat. For field deployment with limited resources, GPU based processing may not be available. For real-time feedback, fast computation on each beat is required.

**Results**

Table 2 provides details for a *subset of hyperparameters and models* tried for the task as well as accuracy achieved on the training or validation set as marked. Figure 5 shows training loss and accuracy on the

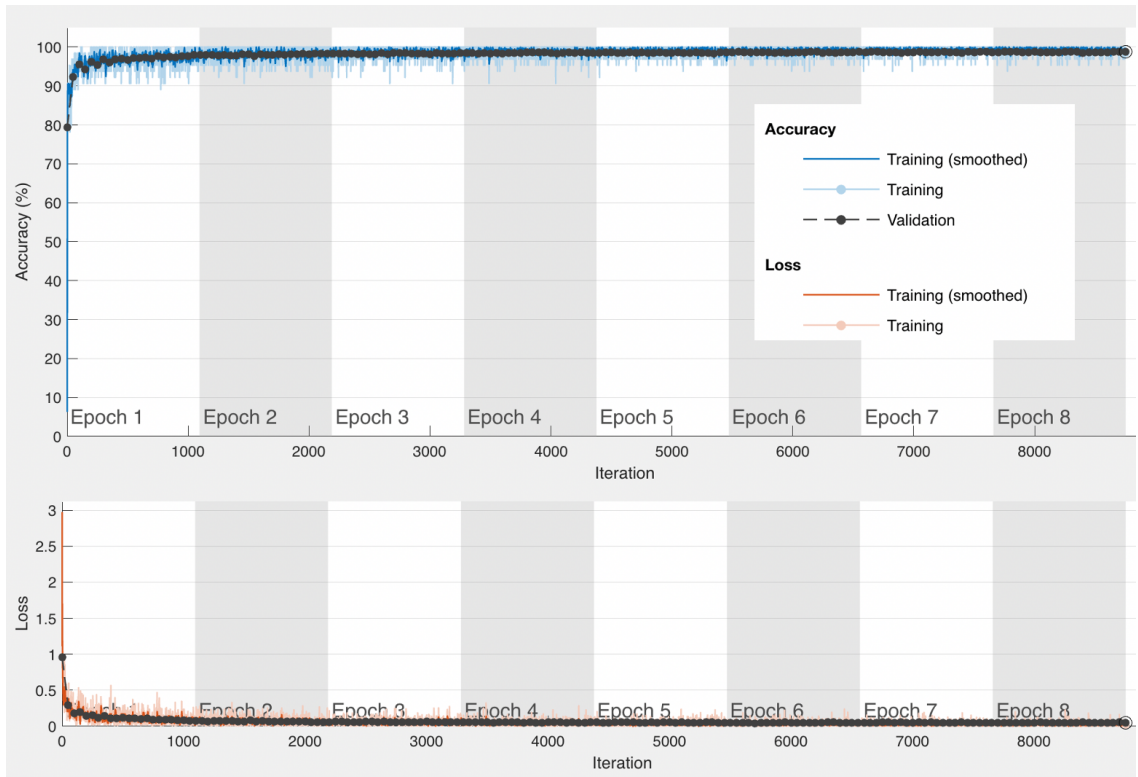

Figure 5: Training and validation accuracy and loss for the ResNet 18 model (schedule 2).

training and validation sets for the ResNet18 based model (schedule 2 in Table 2). Final accuracy achieved on validation set was 98.74%. The precision, recall and F1 scores are shown in Table 3. While F1 score for normal beats was 99.35%, fusion beats had an F1-score of 81.2%, a result of having relatively fewer examples. Average accuracy over the test set was 98.69%. Time to classify one ECG sample was 7.3 ms on

| Schedule | Model | Learning rate | Minimum Δ | Patience | Frozen Layers | Epochs | Accuracy (%) |
|---|---|---|---|---|---|---|---|
| 1 | Baseline | 0.01 | - | - | - | 9 | 97.49 |
| 2 | ResNet 18+ | 0.01 | - | - | None | 8 | 98.74* |
| 3 | ResNet 50 | 0.01 | 0.0005 | 5 | None | 23 | 99.07 |
| 4 | ResNet 50(2) | 0.01 | 0.0005 | 5 | 125 | 24 | 98.55 |
| 5 | ResNet 50(3*) | 0.01 | 0.0005 | 5 | 150 | 22 | 98.68* |
| 6 | ResNet 50(3) | 0.01 | 0.0005 | 5 | 150 | 22 | 99.72 |
| 7 | ResNet 50(3⌘) | Adaptive(Ω) | - | - | 150 | 50 | 99.69 |
| 8 | ResNet 50(3) | Adaptive(Ω) | - | - | 150 | 50 | 99.74 |
| 9 | ResNet 50(3) | Adaptive▼ | - | - | 150 | 30 | 99.89❖ |
| 10 | ResNet 50(3) | Adaptive▼ | - | - | 150 | 30 | 99.96 |

(1) All layers trainable (2) first 125 frozen (3) first 150 frozen; *: 80-20 train-val split, validation accuracy; +: Matlab implementation; 3⌘: (3) with extra Dense(64) layer; Ω: lr = 0.01, 0.5 decay rate/20 epochs; ▼: lr = 0.01, 0.5 decay rate/5 epochs; ❖: augmented data a single CPU (2 GHz Quad-Core Intel Core i5) computer. This would allow for real-time classification of beats which is important in practice.

Table 2: Subset of training schedules carried out for various values of hyperparameters and models to ascertain the best combination for the current task.

|   | precision | recall | F1score | Support |
|---|---|---|---|---|
| F | 87.2340 | 75.9259 | 81.1881 | 162 |
| N | 99.0400 | 99.6523 | 99.3452 | 18118 |
| Q | 99.5641 | 99.4403 | 99.5022 | 1608 |
| S | 93.2059 | 78.9568 | 85.4917 | 556 |
| V | 96.3296 | 96.0635 | 96.1964 | 1448 |

Table 3: Precision (%), recall (%) and F1 score (%) for the test data set using schedule 2.

F1 scores for the five classes and with schedules shown in Table 2 are reported in Table 4. While schedules 6 through 10 showed similar final training accuracy, schedule 10 showed the best performance across all five ECG classes on the test data. Final training loss and accuracy for the optimal schedule (#10) is shown in Figure 6. Training time was 155 s/epoch. F1 scores for the optimal schedule and the five classes is shown in Table 5. In all cases, classes with the fewest training examples (F and S) showed the worst performance scores, both achieving F1 scores around the mid-80% mark.

| Schedule | F1 score (F) | F1 score (N) | F1 score (Q) | F1 score (S) | F1 score (V) |
| --- | --- | --- | --- | --- | --- |
| 1 | 0.69 | 0.99 | 0.97 | 0.75 | 0.94 |
| 2 | 0.8119 | 0.9935 | 0.9950 | 0.8549 | 0.9620 |
| 3 | 0.7682 | 0.9910 | 0.9869 | 0.7842 | 0.9512 |
| 4 | 0.7343 | 0.9889 | 0.9805 | 0.7787 | 0.9335 |
| 5 | 0.7641 | 0.9893 | 0.9775 | 0.7863 | 0.9434 |
| 6 | 0.8274 | 0.9922 | 0.9887 | 0.8250 | 0.9595 |
| 7 | 0.8457 | 0.9914 | 0.9894 | 0.7938 | 0.9600 |
| 8 | 0.8344 | 0.9925 | 0.9891 | 0.8247 | 0.9589 |
| 9 | 0.7581 | 0.9827 | 0.9845 | 0.6976 | 0.9365 |
| 10 | 0.8431 | 0.9933 | 0.9887 | 0.8478 | 0.9649 |

Table 4: F1 scores for the five classes corresponding to schedules described in Table 3. Schedule 10 provided the best results.

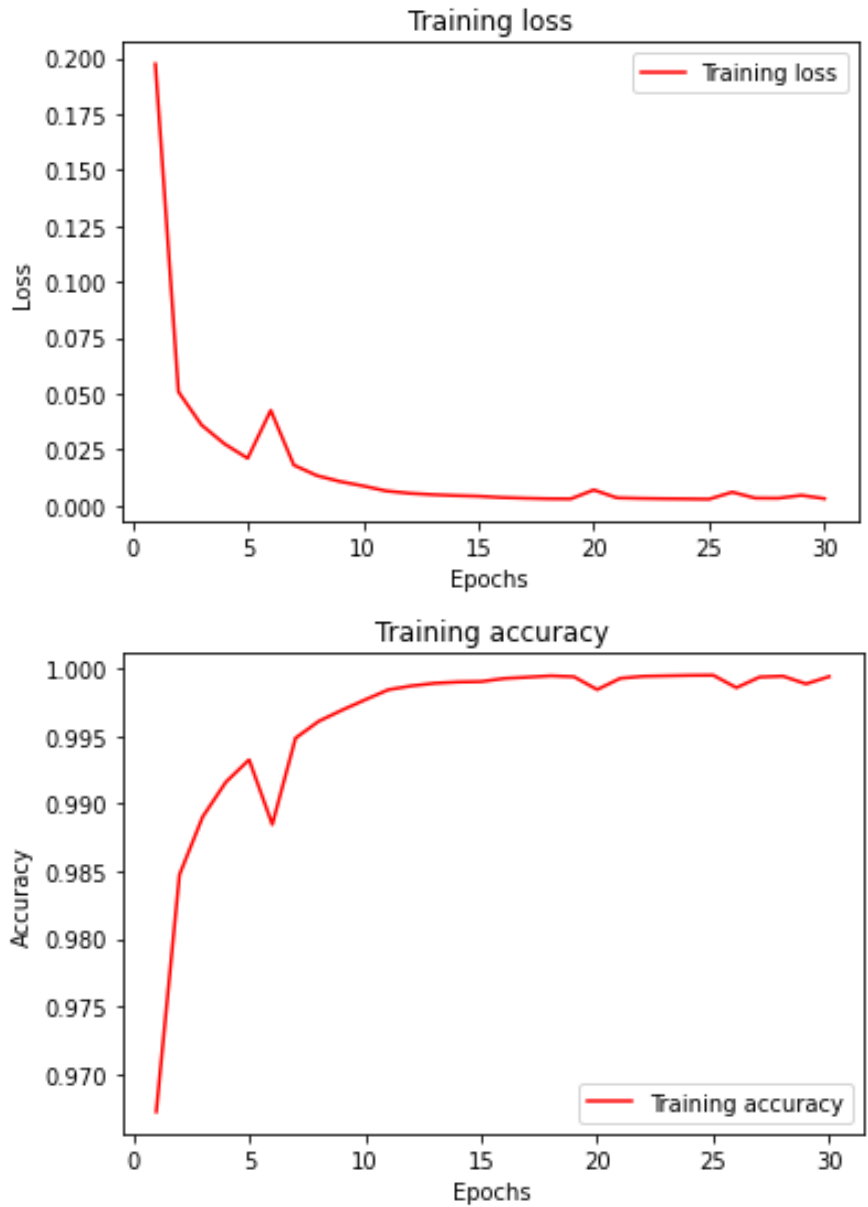

Figure 6: Training loss and accuracy as a function of epochs for schedule 10. As discussed in the text, merging validation data with training data provided the best test set results once hyperparameters were fixed.

As expected, N class had the highest score, followed by Q and V classes, all of which had F1 scores over 95%. Time to classify one beat on the trained saved model was 63 ms on a single CPU (2 GHz Quad-Core Intel Core i5) computer, which would still allow real-time feedback since typical ECG beats are ~ 1 sec in duration. Test results from using WVD images without application of a gradient across the image are shown in Table A1 (Appendix A). Clearly, applying a gradient across the images as done here results in better performance over the test set.

|   | precision | recall | f1-score | support |
|---|---|---|---|---|
| F | 0.8405 | 0.8457 | 0.8431 | 162 |
| N | 0.9905 | 0.9961 | 0.9933 | 18118 |
| Q | 0.9950 | 0.9826 | 0.9887 | 1608 |
| S | 0.9059 | 0.7968 | 0.8478 | 556 |
| V | 0.9706 | 0.9593 | 0.9649 | 1448 |
| accuracy |  |  | 0.9865 | 21892 |
| macro avg | 0.9405 | 0.9161 | 0.9276 | 21892 |
| weighted avg | 0.9862 | 0.9865 | 0.9862 | 21892 |

Table 5: Precision, recall and F1 score for optimal schedule #10. Support refers to number of test examples used for each class.

Figure 7 shows the confusion matrix for schedule 10. The matrix demonstrates that majority of the error in classifying the fusion class arises because fusion arrythmias get confused for normal beats (13/162) and to a lesser extent ventricular arrythmias (12/162). Figure 8 shows examples of F beats and an S beat which were misclassified.

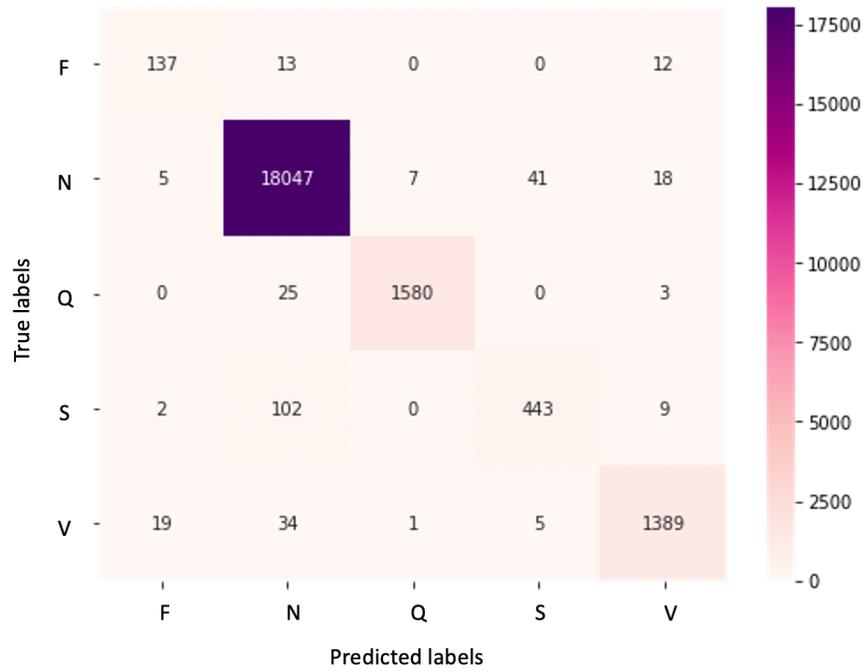

Figure 7: Confusion matrix which captures TN, TP, FP and FN for the 5 classes.

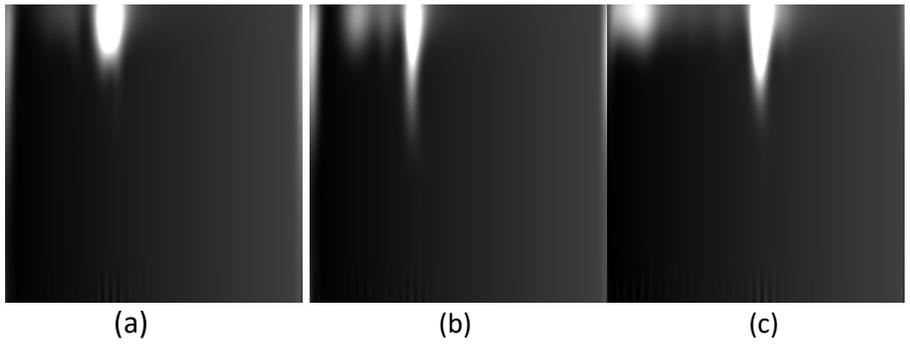

Figure 8: F beats misclassified as a N beat (a) and a V beat (b) while (c) shows a S beat which was misclassified as a N beat.

**Discussion**

Excellent results were achieved after feature extraction and fine tuning a pre-trained ResNet50 model. Early stopping criteria allowed for quicker comparison between different schedules in a relatively short time. Early stopping works well for comparisons but can be sub-optimal once the search is narrowed down to a particular model. This is due to dependency of performance on minimum change in accuracy over patience ($\Delta$ = 0.0005 and 5, respectively) epochs. A different combination of $\Delta$ and patience can yield a better result. Similarly, using all training data without splitting off a validation set provided best results over the test set for all schedules. A larger training set is always desirable and the use of regularization, deeper network, along with decaying learning rate discourages overfitting. Running the model with an adaptive rate over several epochs till asymptotic convergence was obtained, provided optimal results. Running longer than 30 epochs did not result in improved performance for the best model and hyperparameters. Adding LSTM layers to a 1D and 2D CNN model (based on temporal nature of signal) also resulted in deterioration of test set performance.

Data augmentation using added noise resulted in sub-optimal performance, not just for the minority data sets (F, Q, S, V), but also the majority set (N). This is because the majority set was culled down from 72471 examples to 20000 examples using randomly selected samples. The minority sets exhibited poorer performance compared with the un-augmented set due to noise possibly altering characteristics of the underlying waveforms. Simple repetition of waveforms to expand the minority sets also did not improve test performance on the minority sets. It is still possible that augmentation might provide better results with some heretofore to be tried scheme. Another area of exploration would be to include data from several leads instead of lead II (which is the most relevant lead signal) but would result in a much larger data set. It is possible that data from other leads could add information that may allow better discrimination. It is important to note that schemes used in available literature to augment the minority sets have been met with little to no improvement. For example, one implementation available on Kaggle [17] provided a F1 score of 70% and 74% for F and S classes with their augmentation method but separate analysis undertaken provided F1 scores of 75% and 85% with the same un-augmented ECG waveform set as used in this work. Another work [10] used augmentation on the test set but did not provide details about training set.

Another issue with the classification of ECG waveforms based on standard data sets such as the MIT-BIH data set is the imbalance in examples for the 5 classes which resulted in relatively poorer performance over the minority F and S classes. As noted earlier, majority of the error in classifying fusion waveforms arises because fusion arrythmias get confused for normal beats and to a lesser extent ventricular arrythmias. This makes sense since N and V classes dominate the examples, although interestingly the Q (unknown) etiology class is not represented among the false negatives. This could be because some F beats might have characteristic closer to N and V beats rather than Q beats. In the case of supraventricular arrhythmias, most of the misclassified S beats were predicted as N beats. Visual observation shows that various beats exhibit variations which can make it difficult for any model to discriminate between cases that overlap considerably. Reducing the majority sets to match the minority set (training with 641 examples for all classes), reduces the performance across all classes. The one obvious solution would be including more independent examples for F and S arrythmias.

Comparison with earlier works is difficult due to lack of standardization of training and test data in many cases. Kachuee et al [13] tried to address this issue by creating a standardized data set on Kaggle. Based on this set, after data augmentation, they reached an average accuracy of 93.4% on the test data. Individual class metrics on the MIT-BIH ECG data were not reported. As noted above, using an implementation available on Kaggle [17], F1 scores of 75% and 85% for the minority F and S classes, respectively, were derived. A recent work [6] based on the data did not report results on a separate test data set, but merely on the entire training set. Another standardized splitting of the MIT-BIH data set between training and test sets was suggested by de Chazal et al [12] earlier and several works relied on this data set. For example, Alfaras et al. [18] utilized half the recordings for training and half for testing based on [12]. Unknown beats were discarded in the work and only ventricular ectopic beat (VEB) performance was compared across several works. This work reported a recall of 92.7% and precision of 86.1% with lead II data, which superseded the performance of other prior works using that data set. In comparison, the network here achieved recall of 95.9% and precision of 97.1% for VEB, albeit on a different split of the data set. Despite dissimilar data sets, our network also far exceeded the performance reported by works based on the de Chazal data set even after discarding the Q waveforms (Appendix A, Table A2). However, removing the Q set resulted in slightly sub-optimal numbers when compared with data set including the Q set. It should also be noted that the model presented here easily outperforms interpretations of ECG waveforms done by well-trained cardiologists estimated to be 74.9% [19] from meta-analysis of 78 studies. (Compare this to the 98.69% weighted average F1 score achieved here.)

Although the test performance on minority F and S sets gave sub-90% F1 scores (84.3% and 84.8%), it is important to note only 12% of patients with supraventricular arrythmia go on to develop a more serious condition like atrial fibrillation [20]. Fusion arrhythmias are uncommon and can be remedied by pacemaker adjustment, if a fusion of paced and normal beats. On the other hand, ventricular arrythmias have the highest fatality and the model presented provided an F1 score of 96.5% for this important class.

In conclusion, a ML model based on a unique transformation of the ECG waveform along with applying a gradient across the transformed images, followed by design of an optimal deep neural network, provided classification performance that exceeded prior available results from other works.

**Appendix A**

```
              precision    recall  f1-score   support

         0.0     0.8221    0.8272    0.8246       162
         1.0     0.9897    0.9959    0.9928     18118
         2.0     0.9969    0.9845    0.9906      1608
         3.0     0.8932    0.7824    0.8341       556
         4.0     0.9670    0.9510    0.9589      1448

    accuracy                         0.9854     21892
   macro avg     0.9338    0.9082    0.9202     21892
weighted avg     0.9851    0.9854    0.9851     21892
```

Table A1: Training and testing based on WVD images without gradient introduced across images. Training accuracy was 99.97%. However, test accuracy was reduced when compared with images with gradient introduced (Table 5).

```
              precision    recall  f1-score   support

           F     0.8562    0.8086    0.8317       162
           N     0.9906    0.9965    0.9935     18118
           S     0.8956    0.7716    0.8290       556
           V     0.9712    0.9558    0.9635      1448

    accuracy                         0.9859     20284
   macro avg     0.9284    0.8831    0.9044     20284
weighted avg     0.9855    0.9859    0.9856     20284
```

Table A2: Training and testing based on only 4 classes after discarding ECG of unknown etiology (Q) as in [18]. Training accuracy after 30 epochs was 99.87% but accuracy on test data set was sub-optimal when compared with the full data set including Q waveforms.